\let\pdfoutput\defined
\documentclass{PoS}

\PoS{PoS(LAT2005)244}

\title{Properties of phase transitions of higher order\thanks{
Work partially supported by the EU RTN-Network `ENRAGE': {\em Random Geometry
and Random Matrices: From Quantum Gravity to Econophysics\/} under grant
No.~MRTN-CT-2004-005616.}}

\ShortTitle{Properties of phase transitions of higher order}

\author{W. Janke
\\
 Institut f\"ur Theoretische Physik, Universit\"at Leipzig, Augustusplatz 10/11, 04109 Leipzig, Germany\\
E-mail: \email{Wolfhard.Janke@itp.uni-leipzig.de}
}

\author{D.A. Johnston
\\
 Department of Mathematics, School of Mathematical and Computer Sciences, Heriot-Watt University, Riccarton, 
 Edinburgh, EH14 4AS, Scotland 
\\
E-mail: \email{D.A.Johnston@ma.hw.ac.uk}
}

\author{\speaker{R. Kenna}\\
Department of Mathematical Sciences, Coventry University, Coventry, CV1 5FB, England\\
E-mail: \email{r.kenna@coventry.ac.uk}}

\abstract{There is only limited experimental evidence for the existence in nature
of phase transitions of Ehrenfest order greater than two.
However, there is no physical reason for their non-existence, and such
transitions
certainly exist in a number of theoretical models in statistical physics and
lattice field theory.
Here, higher-order transitions are analysed through the medium of partition
function zeros. Results concerning the distributions of zeros are derived
as are scaling relations between some of the critical exponents.}

\FullConference{XXIIIrd International Symposium on Lattice Field Theory\\
25-30 July 2005\\
Trinity College, Dublin, Ireland}

\begin{document}

\section{Introduction}

In the original Ehrenfest scheme for their classification, the order of phase transitions
was given as that of the lowest derivative of the Helmholtz free energy in which a 
discontinuity is manifest. While first-order solid-liquid-vapour transitions and second-order superconducting 
transitions, for example, fit into this scheme, there is only limited evidence for 
the existence in nature of transitions of strictly higher order (i.e., of order three or above).
However, on the basis of experiment,
recent claims have been made which support the existence of a fourth-order transition in a 
cubic superconductor \cite{KuDo99} and a theoretical analysis of 
higher-order transitions was provided in
\cite{KuSa02}.

There are many other well known transitions signaled by divergent
rather than discontinuous behaviour of the appropriate free energy derivatives. 
By far the most common transitions of this type are 
second-order ones such those associated with symmetry breaking of the Higgs field in 
particle physics and ferromagnetic
transitions in metals. The modern classification scheme includes these scenarios and the order 
of a transition is now generally taken to mean that of the lowest derivative of free energy 
in which any type of singular behaviour occurs. 

Notwithstanding the dearth of experimental evidence, higher-order transitions (especially of 
order three) abound in theoretical models, such as pure spin models \cite{spin}, spin models
coupled to quantum gravity \cite{fat1,fat2} and lattice (as well as continuum) toy models of QCD
\cite{QCD}. In the notation of spin models, where $t=1-T/T_c$ is the reduced temperature ($T_c$
being the critical value of the temperature variable $T$) and $h = H/k_BT$ is the reduced external
field ($H$ being the field strength), the critical point is often located at $(0,0)$. This may be the 
endpoint of a line of first-order transitions, in which case we assume that $t$ parameterises its arc length
with $h$ in the orthogonal direction. We denote the free energy by $f(t,h)$
and  its $n^{\rm{th}}$ derivatives by $f^{(n)}_t(t,h)$ and $f^{(n)}_h(t,h)$. If $t$ or
$h$ vanishes, we simply drop it from the argument. 
For an $m^{\rm{th}}$-order transition  $f^{(n)}_t(t)$ is continuous for $n \le m-1$ 
while $f^{(m)}_t(t)$ experiences either a disontinuity or is unbounded at $t=0$. 
It is also possible that $f^{(n)}_h(t)$ is continuous for $n \le m^\prime-1$ 
while $f^{(m^\prime)}_h(t)$ is singular. 
The specific heat,  $ C(t) =  f^{(2)}_t (t)$,
is thus continuous if $ m >2 $  and the susceptibility, $ \chi (t) =  f^{(2)}_h (t)$,
is continuous if $ m^\prime > 2$ (a situation not normally possible in a ferromagnet).

We distinguish between transitions characterised by a discontinuity in $ f_t^{(m)} (t)$
and those characterised by divergences. In the latter case, allowing for the
possibility that $m^\prime$ is not necessarily the same as $m$, 
the scaling behaviour at the transition may be 
described by  critical exponents:
\begin{equation}
 f^{(m)}_t(t) \sim t^{-A} \;, \quad
 f^{(m^\prime)}_h(t) \sim t^{-G} \;, \quad
 f^{(1)}_h(t) \sim t^{\beta} \;, \quad
 f^{(1)}_h(h) \sim h^{1/\delta} \;.
 \label{3}
\end{equation}
In the second-order case ($m=m^\prime=2$), the exponents 
$A$ and $G$ become the usual critical exponents 
$\alpha$ and $\gamma$ associated with specific heat and 
susceptibility, respectively.

Here, some results concerning the properties of partition function zeros close
to higher-order transitions are presented. The Fisher zeros of systems governed by a 
single thermal parameter are analysed in Sec.~\ref{sec:fisher} 
where restrictions on their properties are outlined. 
The Lee-Yang zeros appropriate to the field variable are examined in Sec.~\ref{sec:LY} 
and a number of scaling relations are derived. 
These recover the higher-order relations of \cite{KuSa02} in the case where $m^\prime = m$ 
and, when $m = 2$, they recover the standard Rushbrooke and Griffiths scaling relations
appropriate to second-order transitions where the specific heat and susceptibility are unbounded.
Conclusions are presented in Sec.~\ref{sec:ccl}.

\section{Fisher zeros}
\label{sec:fisher}
\setcounter{equation}{0}

In the situation where the locus of Fisher zeros is linear and
can be parameterized  near the transition point in the upper half-plane
by  
$ t(r) =   r \exp(i \phi(r)) $
-- the locus in the lower half-plane given by it complex conjugate --
the (reduced) free energy  and its derivatives are   
\begin{eqnarray}
 f(t)  =  2 {\rm{Re}} \int_0^R{\ln{\left( t-t(r) \right)}} g(r) dr \; ,
\quad
 f^{(n)}_t (t) 
  = 
 2 (-1)^{n-1}(n-1)!
   {\rm{Re}} \int_0^R{\frac{g(r)}{\left(t-t(r)\right)^n}}  dr 
\; ,
\label{2}
\end{eqnarray}
where $R$ is a suitable cutoff.

The difference in the free energy on either side of 
the transition can be expanded as 
$
 f(t>0) - f(t<0) 
 =
 \sum_{n=1}^\infty{c_n}t^n
$.
For a discontinuous transition, $c_n =0$ for
$n<m$, while $c_m \ne 0$ and the discontinuity in 
the $m^{\rm{th}}$ derivative of the free energy is
$
\Delta f^{(m)} = m!c_m
$.
Matching the real parts of the free energies across the singular line gives 
$
 \sum_{n=m}^\infty{c_n}r^n \cos{n\phi (r)}
 = 0
$,
so that the impact angle  is 
\begin{equation}
 \lim_{r \rightarrow 0}{\phi(r)} \equiv   \phi = \frac{(2l+1)\pi}{2m}
\quad \quad {\rm{for}} \quad l = 0,\dots, m-1
\; .
\label{general}
\end{equation}
The formula (\ref{general}) gives a very direct way to read off
the order of a transition. A list of permitted impact angles in the upper half-plane
for  discontinuous transitions
of various orders is given in Table~1. 
\begin{table}
\caption{Impact angles (in the upper half-plane) 
permitted at a discontinuous phase transition of order $m$.}  
\begin{center}
  \begin{tabular}{|c|ccccccccccccccccc|} 
    \hline
     $m$ &  \multicolumn{17}{c|}{Permitted values of impact angle, $\phi$}  \\
       &   &  &  &  &  & & & &&&&& && &&\\
    \hline
       &   &  &  &  &  & & & & &&&&&& &&\\
     1 &   &&   & & &  &  &   &$\frac{\pi}{2}$&& &  & &&&&\\
     2 &   &&   &  &    $\frac{\pi}{4}$ && & &   &    &  &   & $\frac{3\pi}{4}$&&&&\\
     3 &  & & &  $\frac{\pi}{6}$  && &  &  &$\frac{\pi}{2}$  &    & & &  &$\frac{5\pi}{6}$&&&\\
     4 & &&$\frac{\pi}{8}$   &   &&& $\frac{3\pi}{8}$&  &&  &$\frac{5\pi}{8}$ &  & &&  $\frac{7\pi}{8}$&&\\
     5 && $\frac{\pi}{10}$& & &  &$\frac{3\pi}{10}$&   && $\frac{\pi}{2}$&  & & $\frac{7\pi}{10}$& 
&  & & $\frac{9\pi}{10}$&\\
     6 & $\frac{\pi}{12}$&& & & $\frac{\pi}{4}$ &&  & $\frac{5\pi}{12}$&  & $\frac{7\pi}{12}$ & & 
& $\frac{3\pi}{4}$
&  & & &$\frac{11\pi}{12}$\\
       &   &  &  &  &  & & & &&&&& && &&\\
    \hline      
    \hline     
  \end{tabular}
\end{center}
\end{table}
It is clear that while vertical impact is allowed at any discontinuous
transition of odd order, an impact angle of $\pi/6$, for example is only allowed
at a transition of order $3$ or $9$ or $15$, etc. On the other hand,
a discontinuous second-order transition
with impact angle $\pi/2$ is, for example, not permitted here.
These observations are in accord with various results for first-, second- and third-order
transitions in \cite{LY}, \cite{BlEv} and \cite{fat1,fat2} which are associated 
with  impact angles $\pi/2$ (corresponding to $l=0$),  $\pi/4$ ($l=0$)
and $\pi/2$ ($l=1$), respectively. 
The question as to the mechanism by which the system selects its
$l$-value is open.

Assuming analytic behaviour for the density of zeros, it is straightforward to show that
the necessary and sufficient condition to generate an $m^{\rm{th}}$-order discontinous transition is
\begin{equation}
 g(r) = g_0  r^{m-1} 
\; ,
\label{hjjjj}
\end{equation}
where $g_0$ is a constant. 
Inserting this into (\ref{2}), and using  (\ref{general}), one finds
\begin{equation}
 \Delta f^{(m)} 
 =
 2 \pi g_0 (m-1)! \sin{m\phi}
 =
 (-1)^l (m-1)! 2 \pi g_0
\; ,
\label{318}
\end{equation}
 which generalizes a well known result in the $m=1$ case \cite{LY}.

We next consider an $m^{\rm{th}}$-order diverging transition characterised by (\ref{3}).
To achieve the appropriate divergence for $f^{(m)}_t(t)$ it is necessary that
\begin{equation}
 g(r) = g_0 r^{m-1-A}
\; .
\label{fo}
\end{equation}
Putting this into (\ref{2}), one finds that $f^{(n)}_t(t)$ is continuous for $n<m$, while
\begin{equation}
f^{(m)}_t(t)
=
 - 2 g_0 |t|^{-A} \Gamma (m-A) \Gamma (A)
 \times
 \left\{
 \begin{array}{ll}
        \cos{(m-A)\phi} & \mbox{if $t<0$} \\
        \cos{\left((m-A)\phi+A \pi \right)}  & \mbox{if $t>0$}\;. 
 \end{array}
\right.
\label{end}
\end{equation}
This general relationship between critical amplitudes and impact angles recovers results 
derived in \cite{AbeF} in the $m=2$ case. If $A= 0$ it turns out that $f^{(m)}(t)$
  becomes logarithmically divergent and
\begin{equation}
f^{(m)}_t(t)
=
 2 g_0 (m-1)!\cos{(m\phi)}
 \times
 \left\{
 \begin{array}{ll}
        \ln{|t|}  & \mbox{if $t<0$} \\
        (\ln{|t|} + \pi \sin{(m\phi)} )  & \mbox{if $t>0$}\;. 
 \end{array}
\right.
\end{equation}
When $m=2$ this recovers known behaviour in the second-order case \cite{AbeF}
and the discontinuity across the 
transition is consistent with (\ref{318}).

A quite natural origin for finite-size scaling (which is alternatively and traditionally 
viewed as a hypothesis)
was offered in \cite{JaJo01} by matching the integrated density of zeros on a finite lattice
to its infinite-volume counterpart. Here, from
(\ref{hjjjj}) and (\ref{fo}), this quantity is
$
 G(r) \sim r^{m-A}
$
(where $A=0$ in the case of a discontinuous transition).
For a finite system of linear extent $L$, the integrated density 
at the $j^{\rm{th}}$ zero is
$ G_L(|t_j|) = (2j-1)/2L^d$ \cite{JaJo01,JaJo04}, and equating this
to $G(|t_j|)$  leads to the scaling behaviour
\begin{equation}
 |t_j| \sim L^{-\frac{d}{m-A}}
\; .
\label{FSSF}
\end{equation}
In the unbounded case, assuming hyperscaling ($m-A=2-\alpha = \nu d$), this recovers the standard form
$|t_j| \sim L^{-1/\nu}$ for Fisher zeros.
In the discontinuous case, where $A=0$, (\ref{FSSF}) yields
\begin{equation}
 \nu = \frac{m}{d}
\; ,
\end{equation}
which is a generalization of the usual formal identification of $\nu$ with $1/d$, 
applicable at a first-order transition. 
Support for this identification comes from the third-order ($m=3$) 
discontinuous transition 
present in the $d=3$ spherical model \cite{BeKa52} as well as in the Ising model on 
planar random graphs if the Hausdorff dimension is used for $d$
\cite{fat1}.

\section{Lee-Yang zeros}
\label{sec:LY}
\setcounter{equation}{0}

While the Fisher-zero analysis concerns only even exponents, mixed exponents 
are encountered in the Lee-Yang case, where both field and thermal variables enter.
Scaling relations between such exponents were derived in \cite{KuSa02} in the 
theoretical case where $m^\prime = m$. These are 
\begin{eqnarray}
 (m-1) A + m \beta + G = m(m-1) \;,
\quad 
G =  \beta \left( (m-1) \delta  -1\right) \;.
\label{Griffiths1}
\end{eqnarray}
If $m=2$, these  become 
the standard Rushbrooke and Griffiths scaling laws, respectively:
\begin{equation}
 \alpha + 2 \beta + \gamma = 2 \;, \quad
\gamma =  \beta (\delta -1) \;.
\label{RG2}
\end{equation}

Now, the  free energy in the presence of an external field is
\begin{equation}
 f(t,h) = 2 {\rm{Re}} \int_{r_{\rm{YL}}(t)}^R{\ln{(h-h(r,t))}g(r,t) dr }
\; ,
\label{g1}
\end{equation}
where $r_{\rm{YL}}(t)$ is the Yang-Lee edge and the locus of  zeros is 
$ h(r,t) = r \exp{(i \phi(r,t))}$.
The ${n}^{\rm{th}}$ field derivative is
\begin{equation}
 f^{({n})}_h(t) = \frac{
                       (-1)^{n-1}2(n-1)!
                      }{
                       r_{\rm{YL}}(t)^{n-1}
                      } 
 {\rm{Re}}
\int_1^{  \frac{R}{ 
                     r_{\rm{YL}} 
                    }
        }{ 
\frac{g(xr_{\rm{YL}},t)}{ \left( {h}/{r_{\rm{YL}} } - x e^{i \phi}\right)^n }
     dx }
\; ,
\label{g4}
\end{equation}
having used $r=xr_{\rm{YL}}(t)$.
We assume  that $r_{\rm{YL}}(t)$ is 
small near $t=0$ so that $R/ r_{\rm{YL}}(t) \rightarrow\infty$ 
and, when $n=m^\prime$ and $h=0$, compare with the limiting scaling behaviour in (\ref{3})
to find 
\begin{equation}
 g(r,t) = t^{-G} r_{\rm{YL}}(t)^{{m^\prime}-1} \Phi{\left(\frac{r}{r_{\rm{YL}}(t)}\right)}
\;,
\label{g5}
\end{equation}
for some function $\Phi$ \cite{AbeLY}. Using $n=1$ in (\ref{g4}) gives, for  the magnetization,
\begin{equation}
 f^{(1)}_h(t,h) = 2 t^{-G} r_{\rm{YL}}(t)^{{m^\prime}-1}  
 {\rm{Re}}
\int_1^\infty{\frac{\Phi(x)}{\frac{h}{r_{\rm{YL}}(t)}-xe^{i\phi}} dx }
 \equiv  t^{-G} r_{\rm{YL}}(t)^{{m^\prime}-1}  
\Psi_\phi{\left(\frac{h}{r_{\rm{YL}}(t)}\right)}
\; .
\label{g6}
\end{equation}
Comparison with (\ref{3}) now gives 
$
\Psi{\left(h/r_{\rm{YL}}(t)\right)}
\sim
\left(h/r_{\rm{YL}(t)}\right)^{{1}/{\delta}}
$.
Eq.~(\ref{g6}) must become $t$-inde\-pend\-ent for small $t$ and, under these circumstances,
the behaviour of the Yang-Lee edge is
\begin{equation}
 r_{\rm{YL}}(t)
 \sim
 t^{\frac{G\delta}{({m^\prime}-1)\delta -1}}
\;.
\label{3edge}
\end{equation}
When ${m^\prime}=2$ and $G=\gamma$, 
this recovers  second-order behaviour.
Inserting (\ref{3edge}) into (\ref{g6}),
and comparing the $h \rightarrow 0$ limit of the resulting expression to  (\ref{3}),
yields the scaling relation
\begin{equation}
 \beta = \frac{G}{({m^\prime}-1)\delta -1}
\; .
\label{sg2}
\end{equation}
When ${m^\prime}=m$, this recovers the Griffiths-type scaling relation
(\ref{Griffiths1}),  derived in \cite{KuSa02}.

From 
(\ref{g5}) and (\ref{3edge}), the integrated density of zeros is
$
 G(r,t)   = 
 t^{G(\delta+1)/(({m^\prime}-1)\delta-1)}
 F\left( {r}/{r_{\rm{YL}}(t)} \right)
$
in which 
$
F(x) = \int_1^x{\Phi(x^\prime)dx^\prime}
$,
enabling (\ref{g1}) to be written as
\begin{equation}
 f(t,h) = 2 {\rm{Re}} \int_{r_{\rm{YL}}(t)}^R{\frac{G(r,t) dr}{he^{-i\phi}-r} } 
=
t^{G\frac{\delta+1}{({m^\prime}-1)\delta-1}}
 {\cal{F}}_\phi{\left(
                         \frac{h}{r_{\rm{YL}}(t)}
                 \right)
                }
\; ,
\label{g11}
\end{equation}
for sufficiently small $t$ and where
$
{\cal{F}}_\phi{\left(
                         w
                 \right)
                }
 =
2 {\rm{Re}}
\int_1^\infty{
 {F(x) }/{(w e^{-i\phi}-x)}dx
 }
$.
Its $m^{\rm{th}}$ temperature derivative at  zero-field is therefore 
$
 f_t^{(m)}(t)
 \sim
 t^{G(\delta+1)/(({m^\prime}-1)\delta-1)-m}
$, an expression which, compared with (\ref{3}),  yields the scaling relation
\begin{equation}
 A = m - G\frac{\delta+1}{({m^\prime}-1)\delta-1}
\;.
\label{sg1}
\end{equation}
In the case ${m^\prime}=m$,  (\ref{sg2}) and (\ref{sg1}) recover all four scaling relations
derived in \cite{KuSa02}, and, in the second-order case ($m=2$), they recover
the standard Rushbrooke and Griffiths scaling laws (\ref{RG2}).

\section{Conclusions}
\setcounter{equation}{0}
\label{sec:ccl}

The self-consistency of scaling at higher-order transitions has been examined 
using partition function zeros. 
While strong constraints on the linear locii (impact angles) of 
Fisher zeros associated with discontinuous $m^{\rm{th}}$-order transitions
have been established,
the mechanism whereby the system selects from $m$ possible locii
is an open question and merits further study. 
For a transition characterised by an unbounded $m^{\rm{th}}$ moment, 
the impact angle is intimately linked with amplitude ratios. 
In both cases, analysis of the impact angle may provide a stable 
approach to the analysis of phase transition order and strength.
A complementary analysis of Lee-Yang zeros leads to  
scaling relations between even and odd critical exponents at higher-order transitions 
of a general type. 
Finally, a finite-size scaling analysis shows that the conventional formal identification 
of $\nu$ with $1/d$ that holds at first-order transitions
extends to $\nu = m/d$ for discontinuous transitions of $m^{\rm{th}}$ order.


\end{document}